\documentclass[a4paper,11pt]{article}
\usepackage{jheppub} 
\usepackage{lineno}

\usepackage{xspace}
\usepackage{color}
\usepackage{xcolor}
\usepackage{rotating}
\definecolor{light-gray}{gray}{0.8}

\newcommand{\PbPb}{Pb--Pb\xspace}
\newcommand{\AuAu}{Au--Au\xspace}

\newcommand{\TeV}{\ensuremath{\rm TeV}\xspace}
\newcommand{\fm}{\ensuremath{\rm fm}\xspace}

\newcommand{\Dzero}{\ensuremath{{\rm D}^{0}}\xspace}

\newcommand{\Bplus}{\ensuremath{{\rm B}^{+}}\xspace}
\newcommand{\Bminus}{\ensuremath{{\rm B}^{-}}\xspace}
\newcommand{\Bzero}{\ensuremath{{\rm B}^{0}}\xspace}
\newcommand{\AntiBzero}{\ensuremath{\bar{\rm B}^{0}}\xspace}

\newcommand{\Jpsi}{\ensuremath{{\rm J}/\psi}\xspace}
\newcommand{\TwoSpsi}{\ensuremath{\psi(2S)}\xspace}
\newcommand{\Lambdac}{\ensuremath{\Lambda_{\rm c}}\xspace}

\newcommand{\cmnt}[1]{}

\newcommand{\GeVc}{\ensuremath{\mathrm{GeV}/c}}

\newcommand{\drvd}{{\rm d}}
\newcommand{\Fluidum}{Fluid\ensuremath{u}M}
\newcommand{\Trento}{T\textsubscript{R}ENTo}

\newcommand{\RAA}{\ensuremath{R_{\text{AA}}}\xspace}

\newcommand{\pT}{\ensuremath{p_{\text{T}}}\xspace}

\newcommand{\FastReso}{{\tt FastReso}~}

\title{\boldmath Transverse dynamics of charmed hadrons in ultra-relativistic nuclear collisions}

  \author[a]{Anton Andronic,}
    \affiliation[a]{Institut f\"{u}r Kernphysik, Westf\"{a}lische Wilhelms-Universit\"{a}t M\"{u}nster,\\ Wilhelm-Klemm-Str. 9, M\"{u}nster, Germany}
  \author[b,c]{Peter Braun-Munzinger,}
    \affiliation[b]{Research Division and ExtreMe Matter Institute EMMI, GSI Helmholtzzentrum f\"{u}r Schwerionenforschung GmbH, Planckstr. 1, Darmstadt, Germany}
  \affiliation[c]{Physikalisches Institut, Ruprecht-Karls-Universit\"{a}t Heidelberg,\\ Im Neuenheimer Feld 226, Heidelberg, Germany}
  \author[c]{Hjalmar Brun\ss en,}
  \author[c]{Jana Crkovsk\'a,}
  \author[c]{Johanna Stachel,}
\author[d]{Vytautas Vislavicius,}    \affiliation[d]{Lund University, Professorsgatan 1, Lund, Sweden}
  \author[c]{Martin V\"olkl}
\emailAdd{andronic@uni-muenster.de, p.braun-munzinger@gsi.de, brunssen@physi.uni-heidelberg.de, 
stachel@physi.uni-heidelberg.de,
vytautas.vislavicius@hep.lu.se,
voelkl@physi.uni-heidelberg.de
}

\abstract{Transverse momentum \pT spectra and anisotropic flow distributions  are stu\-died for charmonia and charmed hadrons produced in \PbPb collisions and measured  with the ALICE detector at the CERN Large Hadron Collider (LHC). The investigations are performed within the framework of the Statistical Hadronization Model with the transverse dynamics evaluated using predictions from relativistic viscous hydrodynamics as implemented in the computer codes  MUSIC and \Fluidum. With this essentially parameter-free approach good agreement is obtained for \pT spectra in the range \pT $< 10$ GeV/c. The observed wide distribution in \pT of anisotropic flow coefficients v$_2$ and v$_3$ for charmonia is also well reproduced, while their magnitude is generally somewhat over predicted. This finding may be connected to a difference in spatial distribution  between light and charmed hadrons due to a different diffusion of light and heavy quarks in the hot fireball.}

\begin{document}
\maketitle
\flushbottom

\section{Introduction}
\label{sec:intro}

The production and hadronization of charm quarks is of central interest in current quark-gluon plasma (QGP) research~\cite{ALICE:2022wpn}. In particular, hadrons with one or more charm quarks such as charmonia or (multi-)charm hadrons play an increasing role in our quest to quantitatively understand the expansion and hadronization of the QGP fireballs formed in ultra-relativistic nuclear collisions with heavy beams. In addition, the production yields  of such hadrons contain unique information about the degree of deconfinement of charm quarks reached in such collisions~\cite{Andronic:2003zv,Andronic:2017pug,Andronic:2021erx,Gross:2022hyw}. 

The production probabilities of charmed hadrons in central \PbPb collisions are very well described within the framework of the statistical hadronization model (SHMc), see~\cite{Braun-Munzinger:2000csl,Andronic:2017pug} and the description in the following section~\ref{sec:SHMc}. Within this framework alone there is no information about the transverse dynamics and, in particular, the transverse momentum spectra of the produced charm hadrons. Recently, we have shown~\cite{Andronic:2019wva,Andronic:2021erx} that transverse momentum spectra can be well described if one couples hydrodynamical~\cite{Song:2007ux} or hydro-inspired models such as developed within the 'blast wave' approach~\cite{Siemens:1978pb, Schnedermann:1993ws} to the SHMc. Using these models it is assumed that the particles under consideration, i.e. in our case the charm quarks, participate in the collective expansion and flow developed in the QGP fireball as a consequence of the pressure build-up and and pressure gradients in the hot medium. This success implies that a much more direct route to transverse dynamics in the SHMc is to implement the charmed hadrons, e.g. charmonia, directly into one of the now available relativistic hydrodynamics simulation frameworks. In particular, we focus on hydrodynamic modeling of open charm hadrons and charmonia in the framework of MUSIC~\cite{Schenke:2010rr} and of \Fluidum~\cite{Floerchinger:2018pje}. How this is done in detail is described in section~\ref{sec:hydro} below. A key ingredient of the current approach is that the integrated yield for each charm hadron is taken directly from the SHMc calculation, and that, in order to describe the transverse spectral shape, the hydrodynamic evolution is stopped at the QCD phase boundary. Here our explicit assumption is that, for charmed hadrons, not only the yields but also the spectra freeze out at the phase boundary. For light valence flavor hadrons there is, on the other hand, indication that, while hadron yields are frozen, the expansion continues for a couple of fm/c in the now non-equilibrium hadronic phase and that transverse momentum spectra are determined at kinetic freeze out at a somewhat lower effective temperature. At LHC energies this implies that for charmed hadrons the evolution is stopped when the QGP temperature reaches T$_{pc}= 156.5$ MeV. All charmed hadrons are then produced at the same thermal conditions as hadrons with light (u,d,s) valence quarks~\cite{Andronic:2005yp,Andronic:2017pug}, with the one major difference of the introduction of a charm quark fugacity $g_c$~\cite{Braun-Munzinger:2000csl,Andronic:2021erx}. The quantity $g_c$ is introduced since charm quarks are dominantly produced in initial hard collisions, i.e. never reach chemical equilibrium but thermalize in terms of their momentum distributions subsequently in the hot QGP medium. Thus, $g_c$ is not a free parameter but determined from the measured charm quark content of the fireball. Of course, as in all statistical hadronization models, the approach needs information about the full mass spectrum of charmed hadrons and their strong decays, see~\cite{Andronic:2021erx}. 

The main idea pursued in the present manuscript is to study the implications of the detailed hydrodynamic evolution on the transverse dynamics.
In particular, we compare the predictions of the above cited two, in their technical implementation very different, hydrodynamic model codes with the measured transverse momentum spectra of charmonia and charm hadrons. In addition, we investigate how well the recently measured azimuthal anisotropies of charmonia~\cite{ALICE:2013xna,ALICE:2020pvw} and charm hadrons~\cite{ALICE:2013olq} are reproduced. This is a parti\-cu\-lar\-ly interesting question as first attempts in this direction missed the experimental data in particular at higher values of \pT $>$ \, 4 GeV/c~\cite{Zhou:2014kka,Du:2015wha}. In a recent publication~\cite{He:2021zej} good agreement was obtained by employing Langevin simulations to more accurately account for spatial anisotropies and introducing space-momentum correlations of the charm quarks during the hydrodynamic expansion of the hot fireball. Below we show that both transverse momentum spectra  and anisotropic flow distributions are in rather good
agreement with SHMc predictions under the assumption that charm quarks fully participate in the hydrodynamic expansion. This is realized by directly inserting the charmonia when the hydrodynamic evolution is stopped at the pseudo critical temperature of the chiral phase transition such that charmonia inherit the full flow of the medium at this instance. The direct use of the hydrodynamic models rather than the simplification of a blast wave model (as in \cite{Andronic:2021erx}) retains complexities of the medium evolution. For example, while in blast wave models freeze out is realized as a simple integration along a freeze out contour in a plane of proper time vs radius, in a hydrodynamic model individual fluid cells are followed in time and freeze out occurs when a given energy density or temperature is reached. This implies complex geometries of the freeze out in space and time, with velocity vectors of a given fluid cell and freeze out surface normals pointing in all possible directions. Inspecting the radial profile of velocities of fluid cells, one finds indeed a strong correlation (as employed in the blast wave model) but, in particular for larger radii of 5-10 fm, a very wide distribution of velocities \cite{Andronic:2021erx}. We also note that freeze out at early times actually gives a substantial contribution to particle spectra from fluid cells at the surface of the fireball and with a large velocities. In addition, the use of anisotropic modeling (already in 2+1 dimensional hydrodynamics) allows for calculation of the flow coefficients.

In section~\ref{sec::BWMatching} we show how a blast wave parametrization can be optimally matched to reproduce the full hydrodynamics calculation of the \Jpsi spectrum. Finally in section~\ref{sec:charmspacetime} we discuss and estimate the consequence of the charm spatial distribution possibly being more compact than the energy distribution at freeze out.

\section{Brief summary of the statistical hadronization model for heavy quarks, SHMc}
\label{sec:SHMc}

Here we summarize the main ideas  behind the  SHMc.  For much more detail on the original development  see~\cite{Andronic:2021erx,Andronic:2017pug,Andronic:2003zv,Braun-Munzinger:2000csl}. Our main emphasis will be on the production of charmonia and in particular on the connection between SHMc for yields, and hydrodynamic models for the description of transverse dynamics. The key idea is based on the recognition that, contrary to what happens in the (u,d,s) sector, the heavy (mass $\sim$ 1.2 GeV) charm quarks are not thermally produced. Rather, production takes place in initial hard collisions. The produced charm quarks then thermalize in the hot fireball, but the total number of charm quarks is conserved during the evolution of the fireball~\cite{Andronic:2006ky} since charm quark annihilation is very small. In essence, this implies that charm quarks can be treated like impurities. Their thermal description then requires the introduction of a charm fugacity $g_c$~\cite{Braun-Munzinger:2000csl,Andronic:2021erx}. The value of $g_c$ is not a free parameter but experimentally determined by measurement of the total charm cross section, relying on the precisely measured cross section for D$^0$ production in \PbPb collisions \cite{ALICE:2021rxa} and fragmentation as in the SHMc, see the section on quark matter in~\cite{Gross:2022hyw}. For 0-10\% central \PbPb collisions at 5.02 TeV, the number of $\mathrm{c\bar{c}}$ pairs per unit of rapidity is 13.8$\pm$2.1 (16.3$\pm$2.4 with an enhanced charm-baryon spectrum, see below) resulting in $g_c=31.5\pm 4.7$ and in a D$^0$ to \Jpsi ratio of 53, in very good agreement with recent precision data~\cite{ALICE:2023gco}. Note that, with this approach, unlike in the case in which one starts with the charm production cross section in pp collisions, the uncertainties related to gluon shadowing in the Pb nucleus are avoided, leading to reduced uncertainties in $g_c$ (currently $\pm$15\%).

We note here that the charm balance equation should contain canonical corrections for more peripheral collisions or for lighter collision systems (or lower collision energies), i.e., whenever the number of charm pairs is not large compared to 1~\cite{Gorenstein:2000ck,BraunMunzinger:2003zd,Andronic:2021erx}. For central \PbPb collisions at 5.02 TeV the canonical correction factors are in fact very close to 1~\cite{Andronic:2021erx}. For more peripheral collisions or smaller systems the canonical corrections are important. We follow everywhere the canonical treatment described in detail in~\cite{Andronic:2021erx}.

The large value of $g_c \approx  30$  for charm production in central \PbPb collisions 
at mid rapidity implies very large enhancements for charmed hadrons compared to what is obtained in the purely thermal case. In the absence of canonical corrections the enhancement factor is about 900 for charmonia and doubly charmed, and $ 2.6 \cdot 10^4$ for triply charmed hadrons. 

The statistical hadronization approach is applied to the core of the nuclear overlap region. As the density drop off in nuclei is gradual, we define a corona region. We assume this to start where the density of one of the two colliding nuclei has dropped to 10 \% of the central density. The rationale is that in the overlap of this tail of the distribution with the core or tail of the density distribution of the other nucleus, the average number of collisions is less than 1. The two regions are derived from a Glauber model calculation as a function of centrality. As an example, for the 0-10 \% centrality bin, the total overlap region contains 361.3 participants, of which we define 340.6 as the core. In this core there are 1645 binary collisions. 

\section{Extracting momentum information from hydrodynamic calculations}
\label{sec:hydro}

As described in the previous section, the core part of the fireball is assumed to hadronize statistically while preserving the fixed number of charm and anti charm quarks produced in hard initial collisions. This implies that the relative distribution of the charm quarks among hadrons is driven by the statistical weights at a given temperature. We use the temperature determined for chemical freeze out obtained when fitting the yields of hadrons with u,d,s valence quarks, T$_{chem}$ = 156.5 MeV~\cite{Andronic:2017pug}, which happens to coincide with the pseudo critical temperature for the chiral phase transition T$_{pc}$. While the statistical hadronization model as such makes no prediction about the momentum distributions of the produced hadrons, the underlying assumption is nevertheless that the hadronizing charm quarks are in (or close to) local thermal equilibrium kinetically. Measurements of the various D-meson spectra and elliptic flow combined with model comparisons confirm this assumption \cite{ALICE:2021rxa}. Based on this statistical approach of charmed hadron formation a consistent prediction of the transverse dynamics is obtained by using the hydrodynamic modelling optimized for observables of light valence quark hadrons, albeit at T$_{pc}$. Local thermal equilibrium gives the momentum dependent yields in the rest frame of the fluid, i.e. in each fluid cell. Employing the boosts of all fluid cells and their geometric arrangement, transverse momentum spectra including azimuthal asymmetries can be obtained as described in this section.

In hydrodynamic simulations, the transition from the hydrodynamic phase to a hadron gas with little interaction is usually not modeled as part of the system evolution. Instead, the system evolves in time fully hydrodynamically. The end of the evolution is reached by defining a transition hypersurface, delineating the points in space and time where a certain critical temperature or energy density are reached. At this hypersurface particles are produced according to the Cooper-Frye formalism~\cite{Cooper:1974mv}:

\begin{equation}
    E \frac{\drvd N}{\drvd^3 p} = g_i \int_\Sigma f(u^\mu p_\mu) p^\mu \drvd^3 \Sigma_\mu ~. \label{eq::CF}
\end{equation}

This formalism assumes a thermalized momentum distribution $f$~at the freeze out hypersurface and then counts the particles crossing the boundary. We assume that the charmed hadrons are produced at the same rate in all unit volumes at the transition temperature. This means that information about the shape of the overall momentum distribution can be obtained by calculating the freeze out momentum distribution from the hydrodynamic simulations according to \ref{eq::CF}. The integral of the distribution is then scaled according to the SHMc as described above. Compared to a description using a blast wave approach this gives a much more realistic description of the freeze out process. Here, the freeze out has significant contributions from hypersurfaces that are volume like (with time like surface normal vectors $\drvd^3\Sigma_\mu$) and ones that are surface like (with space like normal vectors). In the blast wave approach only volume like contributions are considered. Also, the full complexity of the radius-time-velocity association is reflected in the spectrum, while in the blast wave approach a simplified tight connection between these variables is employed. Further and new information on the physics behind the Cooper-Frye formalism can be found in~\cite{Kirchner:2023fsj}.

For our purposes we use two hydrodynamic models: MUSIC~\cite{Schenke:2010rr} and \Fluidum~\cite{Floerchinger:2018pje}. MUSIC was used with the settings of \cite{Schenke:2020mbo} and using a 2+1D, boost-invariant setup for the measurements at mid rapidity. This includes an initial state based on the IP-Glasma model~\cite{Schenke:2012hg,Schenke:2012wb}. In \Fluidum, the fluid dynamic fields are decomposed into an azimuthally and Bjorken boost symmetric background contribution and perturbations. This allows a numerically very efficient calculation of the hydrodynamic evolution. For the calculation of transverse momentum spectra, terms of linear order in perturbation do not contribute so that in this case the azimuthally symmetric background evolution is sufficient. \Fluidum\, was used with the parameters as in~\cite{Devetak:2019lsk} and a normalization of the energy density in the initial state model \Trento~\cite{Moreland_2015} that reproduced the integrated pion, kaon and proton yields measured in central \PbPb collisions. In a very recent paper, the diffusion of charm quarks and its effect on transverse momentum spectra of charm hadrons is discussed in the context of the \Fluidum~model~\cite{Capellino:2023cxe}.
The settings of both models are summarized in Table \ref{tab:parameters}. In both cases, a constant ratio of shear viscosity to entropy density $\eta/s$ is assumed. The ratio of bulk viscosity to entropy density $\zeta/s$ is assumed to peak somewhat above the pseudo critical temperature and the values and positions of the maximum are given in the table. The precise functions can be found in \cite{Schenke:2020mbo} and \cite{Devetak:2019lsk}, respectively. 

\begin{table}[h]
	\centering
	\caption{MUSIC and \Fluidum~ parameters}
	\label{tab:parameters}
	\begin{tabular}{|llccccc|}
		\hline
		model & initial conditions & $\tau_0$ & $\eta/s$ & $\left(\zeta/s\right)_\text{max}$&$T_\mathrm{peak}$&$T_\text{fo}$\\ 
        \hline
		\Fluidum & \Trento & $0.18\,\mathrm{fm/c}$ & 0.16 & 0.06 & $175\,\mathrm{MeV}$&$156\,\mathrm{MeV}$\\
		\hline
        MUSIC & IP-Glasma & $0.4\,\mathrm{fm/c}$ & 0.12 & 0.13 & $160\,\mathrm{MeV}$& $156\,\mathrm{MeV}$ \\
        \hline
	\end{tabular}
\end{table}

\section{Calculation of \Jpsi spectrum with MUSIC and \Fluidum}
\label{sec:Hydrospectrum}

\begin{figure}[htbp]
\centering %
\includegraphics[width=.48\textwidth]{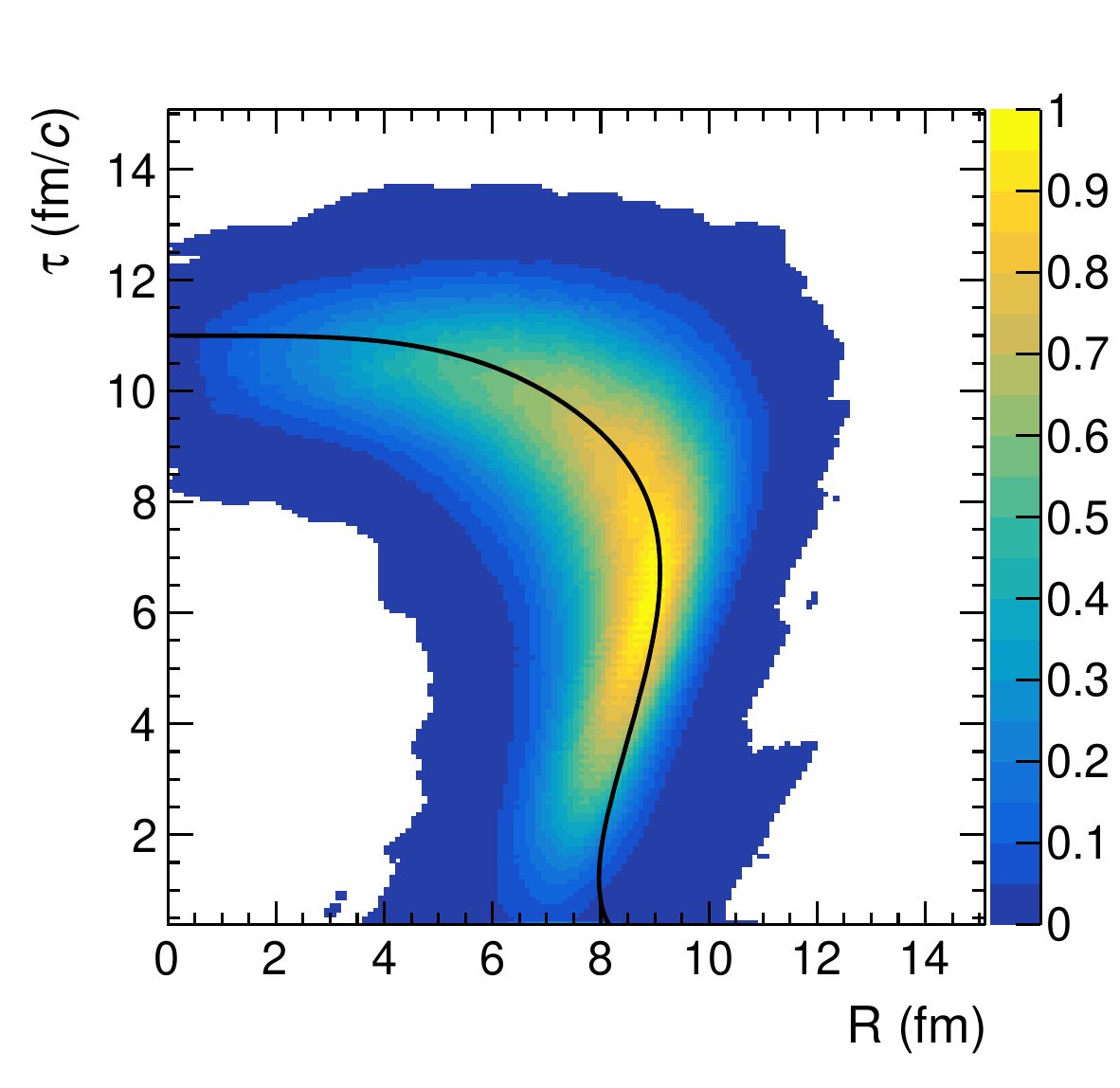}
\includegraphics[width=.48\textwidth]{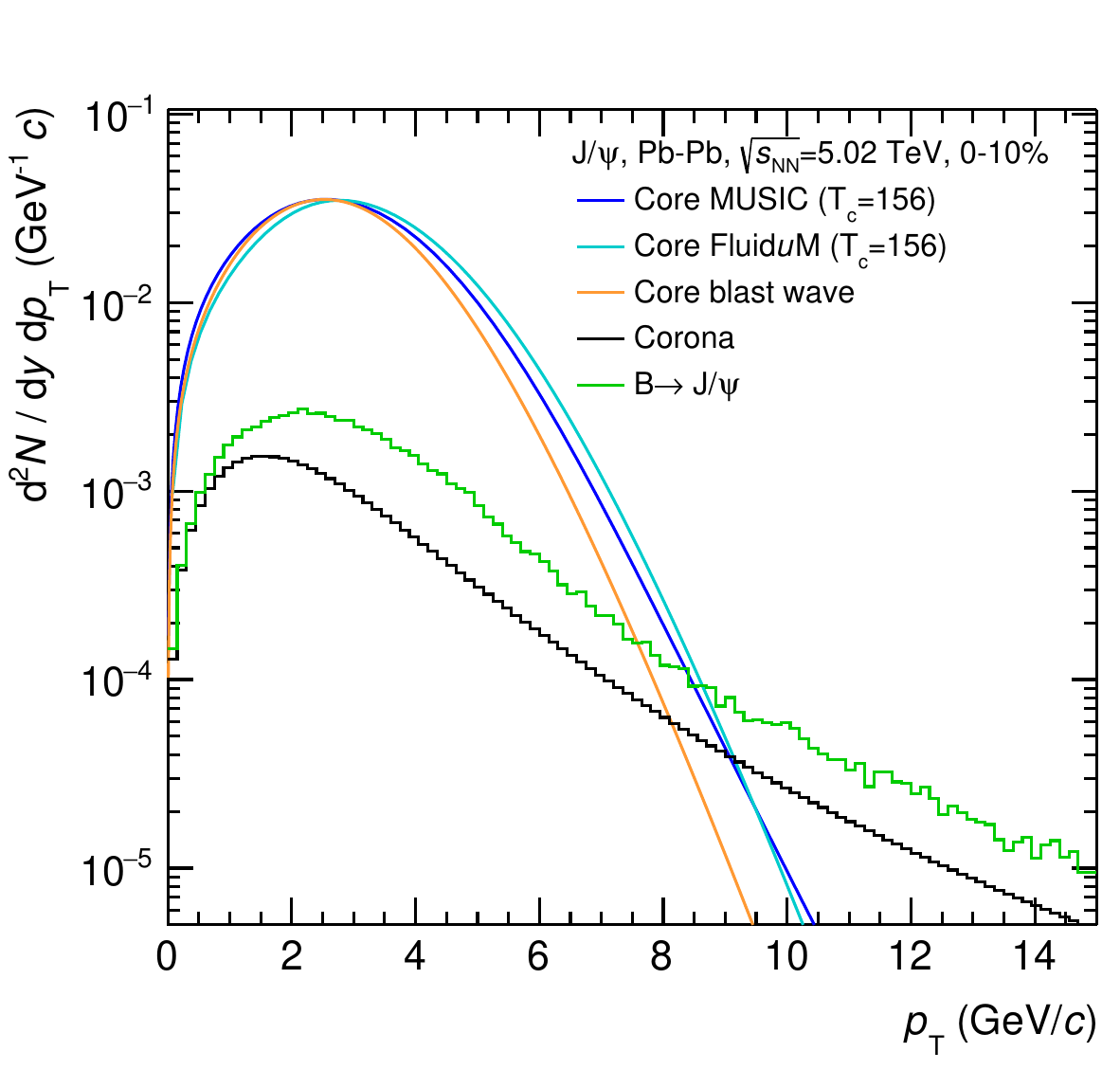}
\caption{Left: $R$ and $\tau$ distribution of the freeze out hypersurfaces for MUSIC (histogram) and \Fluidum~(black line). The MUSIC hypersurface elements are weighted by their volume (but not the energy streaming outwards through this element).
Right: Contributions to the \Jpsi \pT-distribution in the SHMc model. The extraction of the blast wave parameters is discussed in section \ref{sec::BWMatching}. \label{fig:RTau}}
\end{figure}

Figure \ref{fig:RTau} shows the distribution of the freeze out hypersurfaces in time and radius. In \Fluidum, the calculation is started with a rotationally symmetric solution as the basic output around which perturbations are calculated. This means that the resulting freeze out hypersurface is azimuthally isotropic, which gives a line in the $\tau$-$R$ space. 

The \pT\, distribution of thermal \Jpsi mesons resulting from this freeze out represents the contribution of the core in the SHMc model. To this, a pp-like contribution of the corona is added (see above). The third component to the \Jpsi spectrum is obtained from feed down from weak decays of beauty hadrons. This dominantly contributes at high momenta. Beauty quarks interact with the medium, but are likely not fully thermalized~\cite{Andronic:2022ucg}. Thus, the transverse momentum distribution of the beauty hadrons is expected to lie somewhere between the extremes of a pp-like distribution and a fully thermalized distribution with strong effects of collective flow as represented by hydrodynamic simulations. We chose the arithmetic mean of the two cases. For this, we used a pp-like \pT~distribution obtained from FONLL \cite{Cacciari:1998it} and a thermalized distribution represented by blast wave parameters described below. We included B mesons by using PYTHIA8 \cite{Bierlich:2022pfr} to generate their weak decays to \Jpsi. For the sum of \Bplus, \Bminus, \Bzero, and \AntiBzero we used a total yield after all strong decays of $\drvd N/\drvd y=0.86$ and 0.116 for 0-10\% and 30-50\% centrality ranges as given by SHM calculations for the bottom sector at mid rapidity \cite{Andronic:2022ucg}.

\begin{figure}[htbp]
\centering
\includegraphics[width=.48\textwidth]{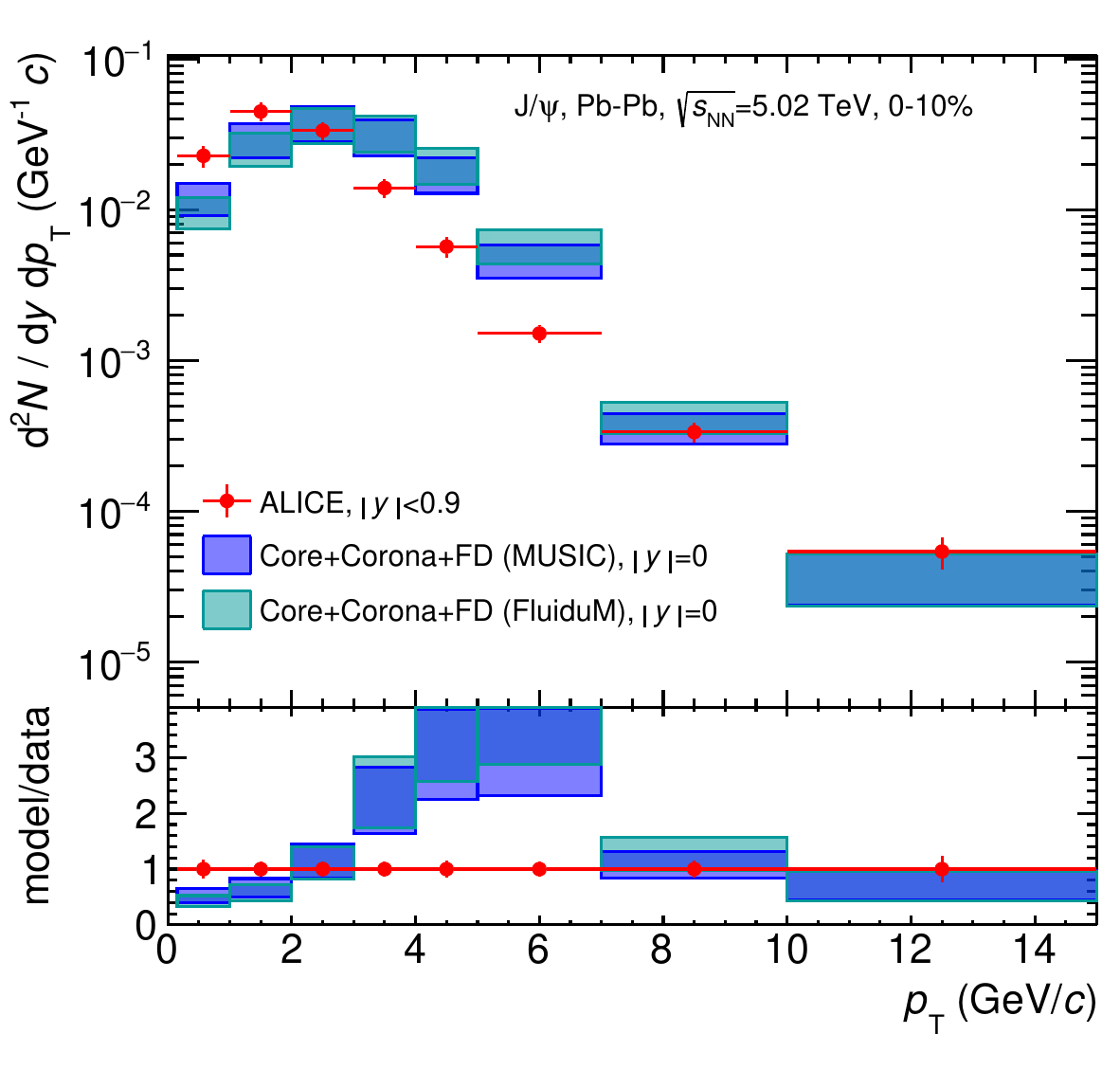}
\includegraphics[width=.48\textwidth]{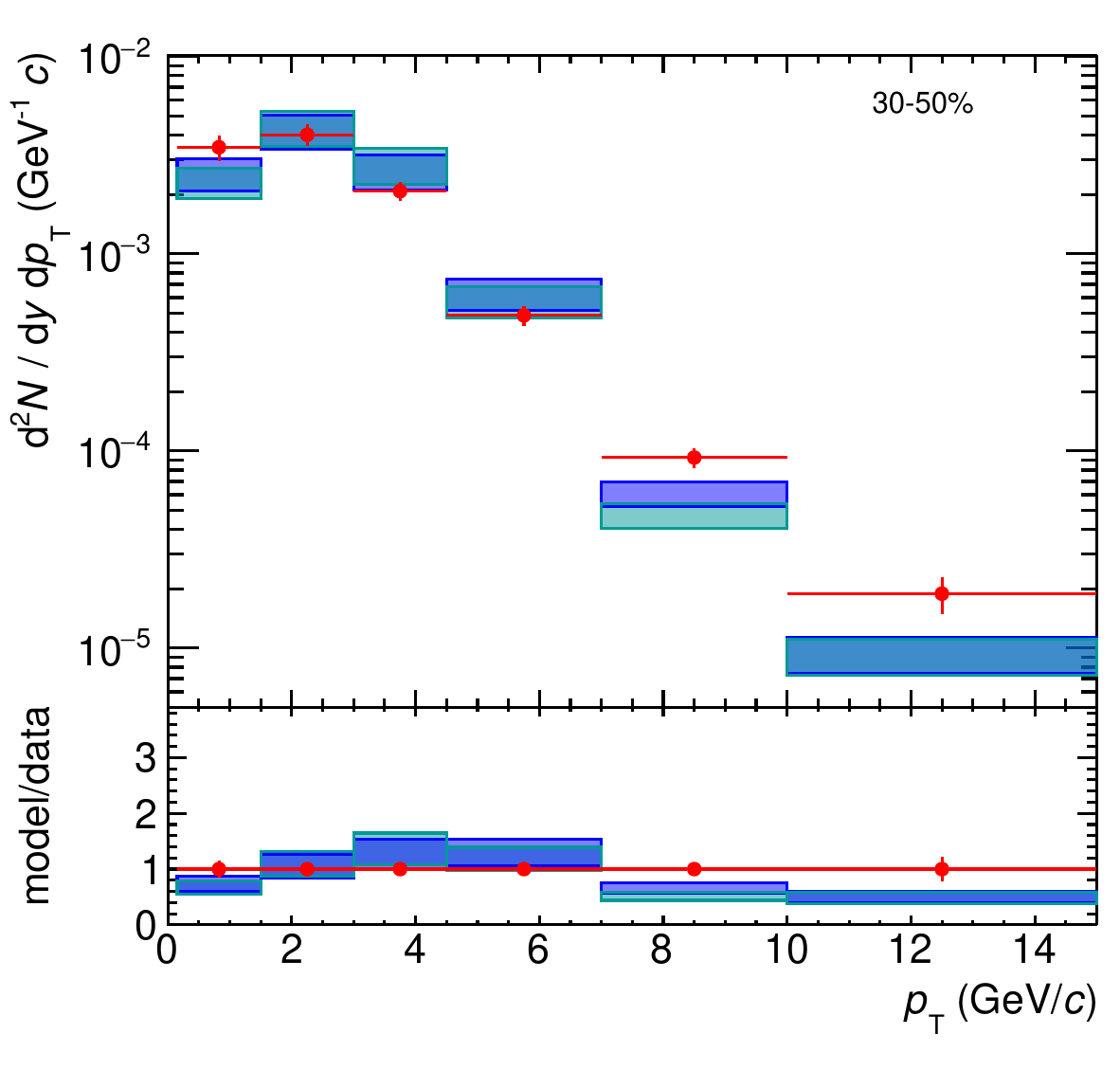}
\caption{Transverse momentum distributions of \Jpsi production in \PbPb collisions at mid rapidity. The result from the SHMc model using the two hydrodynamical generators is compared to the measurement by the ALICE Collaboration~\cite{ALICE:2023gco}. The bottom plot show ratios of model results to the measurements. There the red data points are displayed at unity to indicate the experimental uncertainties. \label{fig:Spectra}}
\end{figure}

Figure \ref{fig:Spectra} shows the resulting \pT~distributions for \Jpsi mesons produced in central and semi-central collisions at mid rapidity together with the ALICE measurement at 5~\TeV. The centrality selection for both models is based on final state particle multiplicities as used in the ALICE experiment. The vertical size of the boxes shown reflects the uncertainty in the charm cross section, i.e. of $g_c$, the uncertainty of the corona contribution due to the experimental uncertainty of the spectrum measured for pp collisions, as well as an uncertainty from the modeling of the feed down from beauty hadrons. The overall agreement is rather good considering that there is no free parameter adjusted to data. The general shape of the distributions is very similar. For the most central bin, the SHMc model gives a distribution that is somewhat harder than the experimental data. The difference is in the intermediate \pT region dominated by the thermalized core contribution from the hydrodynamic simulations, with very similar results from MUSIC and \Fluidum. In the semi-central case, where the transverse flow effect is reduced, the agreement with the data is rather close.

For light hadrons~\cite{ALICE:2019hno}, central as compared to semi-central events result in a larger average \pT~for each particle species. This is well reproduced by an increased mean transverse velocity $\langle\beta_\perp\rangle$ obtained from hydrodynamic models and also from simple blast wave fits to the data. We obtain a corresponding shift in the core contribution using both hydrodynamic generators as seen in Fig. \ref{fig:Spectra}. In the experimental data, however, the maximum of the \pT distribution is, if anything, slightly lower for the more central events. This may point to some additional effect beyond the assumption that the charm quarks are carried along the hydrodynamical flow of the medium.

\begin{figure}[htbp]
\centering
\includegraphics[width=.48\textwidth]{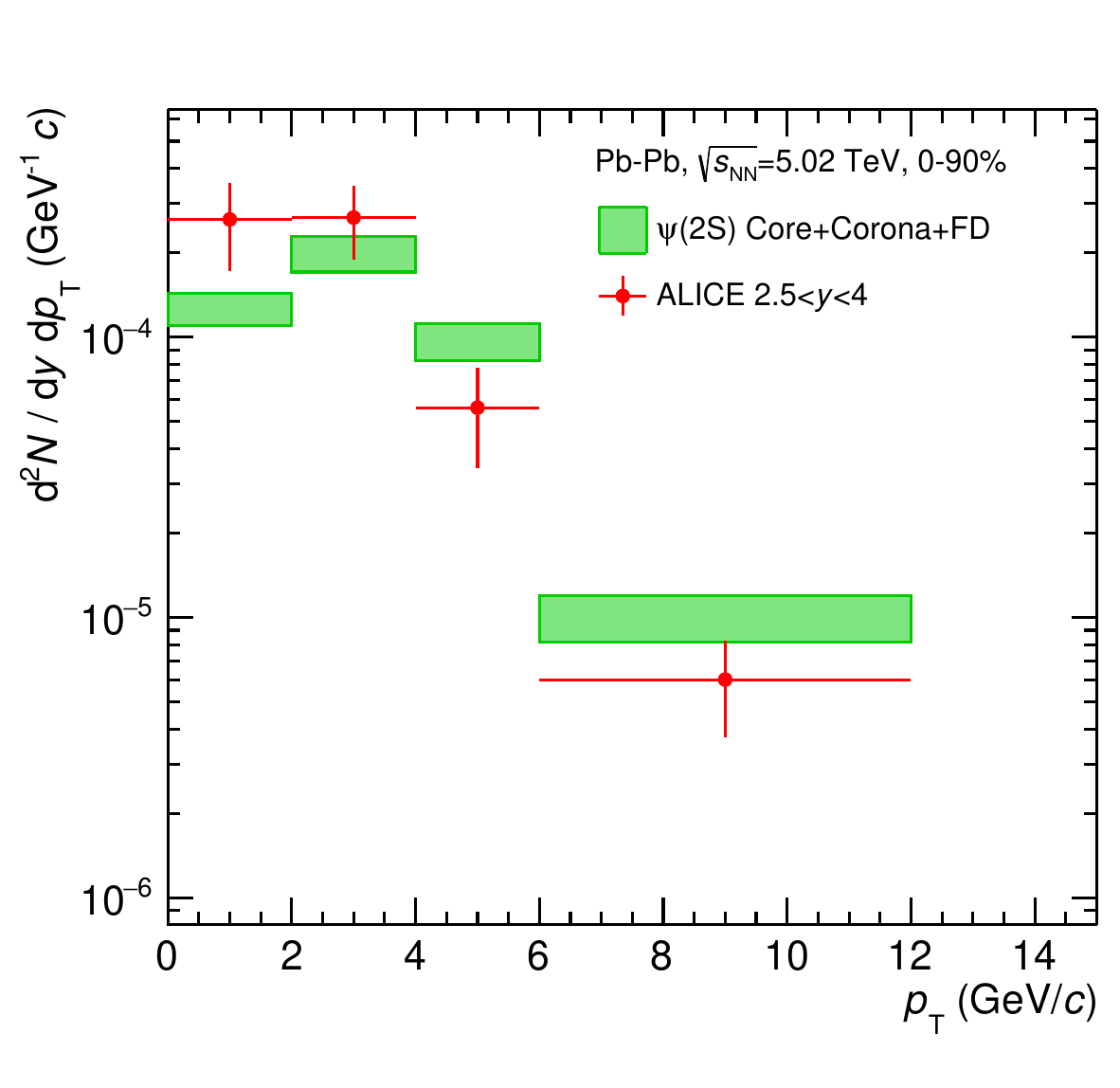}
\caption{Calculation of the \TwoSpsi \pT spectrum in central \PbPb collisions with the SHMc and MUSIC. In comparison are shown experimental data from ALICE \cite{ALICE:2022jeh}. The error band is obtained as for the \Jpsi spectrum.} \label{fig:psi2S}
\end{figure}

In Figure \ref{fig:psi2S} we show the corresponding model prediction for the \TwoSpsi spectrum. The model uncertainties were obtained as for the \Jpsi case above. In comparison we show experimental data from ALICE for \PbPb collisions in the 0-90 \% centrality range \cite{ALICE:2022jeh}.  While the overall spectral shape is in reasonable agreement, the data exceed the model predictions at low \pT by 1-2 standard deviations. High precision data expected from ALICE Run 3 for central collisions are needed to establish whether there are any significant differences between model predictions and data.

Comparing to our first modelling~\cite{Andronic:2019wva} of charmonium spectra using a blast wave model, we note that the current spectra based on full hydrodynamics are somewhat harder. The most probable \pT increases from about 1.5 GeV/$c$ in the blast wave calculations to 2.5 GeV/$c$ with full hydrodynamic treatment, a trend which was in fact expected. The blast wave model parameters were chosen to match the MUSIC distribution of the freeze out hypersurface with a polynomial of the velocity $\beta$ as function of radius. While the MUSIC distribution could be well matched with an exponent of n = 0.85 and a value of $\beta_{max}$ = 0.62, it was apparent that the distribution was very broad for fluid elements at radii between 5 and 10 fm reaching from velocities of about 0.1 to 0.85 and therefore some arbitrariness in how to define the best $\beta_{max}$ (see section \ref{sec::BWMatching} below). 

\section{Modeling of \Jpsi anisotropic flow coefficients}
\label{sec:Hydroflow}

The flow coefficients quantify the azimuthal anisotropy of particle distributions, driven in hydrodynamic models by the azimuthal anisotropy and resulting pressure gradients in the initial state and reduced by dissipation during the expansion phase. In the MUSIC calculations we have full access to the IP-Glasma initial state. We use the definition of the minor axis of the initial state ellipse and triangularity 
\begin{equation}
    \psi_{n} = \frac{{\rm atan2}(r^2 \langle \sin(n\phi) \rangle, r^2 \langle \cos(n\phi) \rangle )+\pi}{n} ~,
\end{equation}
as introduced in \cite{Alver:2010gr}. Here, the average was taken over the distribution of the energy density in the initial state rather than that of the participants used in the original approach. The flow coefficients $v_2$ and $v_3$ were then calculated as

\begin{equation}
    v_n = \langle \cos(n (\phi-\psi_n)) \rangle ~,
\end{equation}
with the averaging done over the expectation values of the number of \Jpsi in each $\phi$ bin. The resulting flow corresponds to fully thermalized charm quarks flowing with the medium and producing \Jpsi mesons during hadronization. This was taken as the flow contribution of the core part of the model. The contribution of the corona was assumed to be independent of the angle $\phi$, consistent with initial production without further interaction. The feed down from beauty hadrons was modeled as having a $v_2$ and $v_3$ constant in \pT over the range shown, with values of $v_2=0.01\pm 0.01$ and $v_3=0.0\pm 0.02$ for the 0-10\% and $v_2=0.05\pm 0.02$ and $v_3=0.0\pm 0.02$ for 30-50\% centrality bins. These values were estimated from the measurements in \cite{ATLAS:2020yxw}.

\begin{figure}[htbp]
\centering
\includegraphics[width=.48\textwidth]{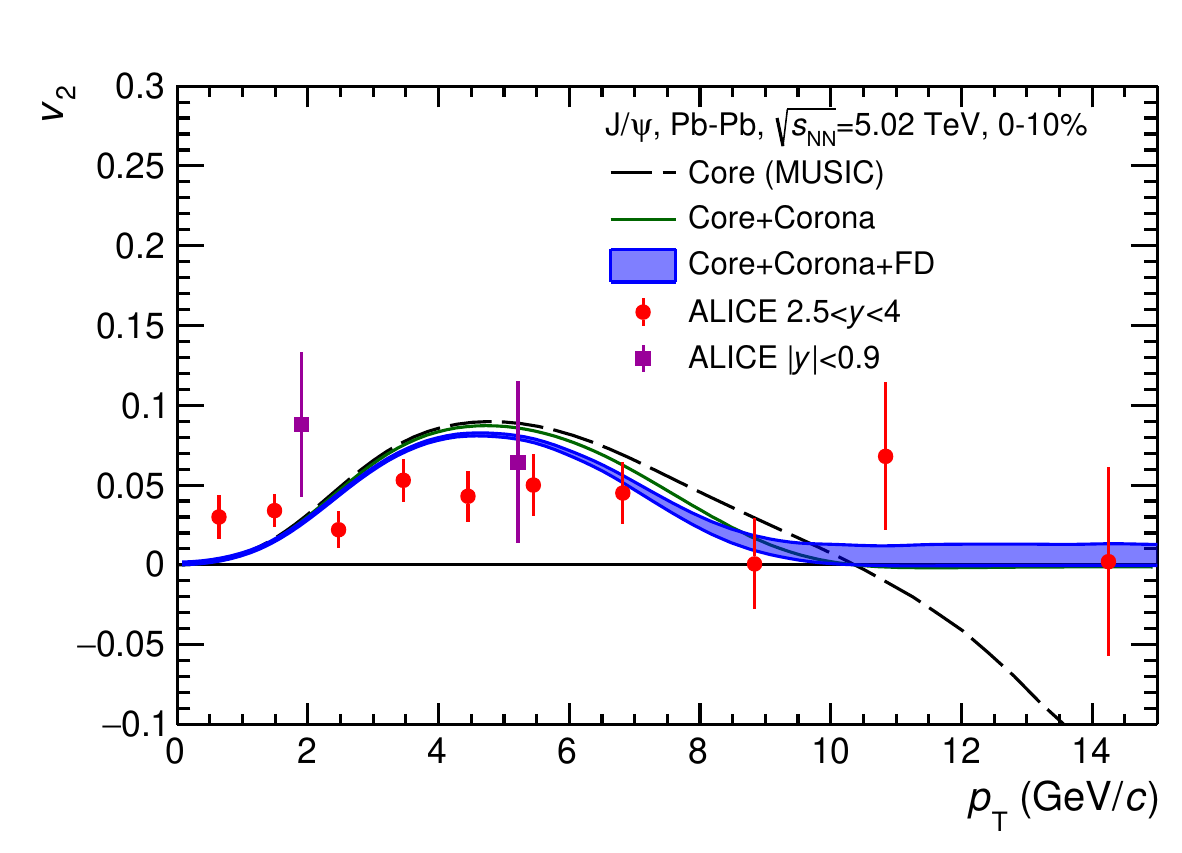}
\includegraphics[width=.48\textwidth]{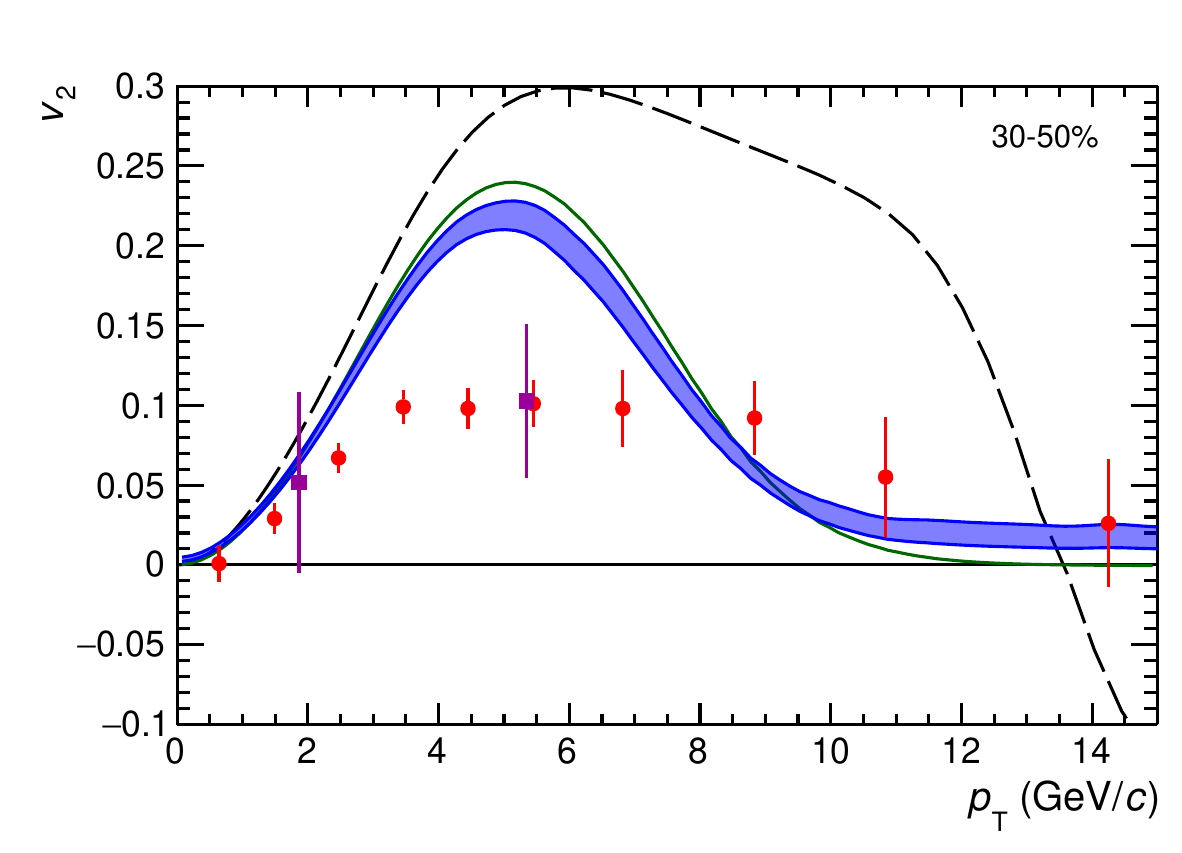}
\caption{The \, $v_2$ coefficients of the SHMc model prediction combined with the 2+1D MUSIC calculations. For the error band see text. The results are compared with data from the ALICE Collaboration at central and forward rapidity for central (left) and semi-central (right) \PbPb collisions~\cite{ALICE:2020pvw}.\label{fig:v2}}
\end{figure}

Figure \ref{fig:v2} shows the resulting \pT-dependent elliptic flow coefficients for \Jpsi in central and semi-central events. The error band contains the uncertainty of the overall charm cross section, the experimental uncertainty of the pp spectrum and the uncertainty attributed to the beauty feed down contribution. The elliptic flow at large \pT~is suppressed by the increasing contribution from the corona. This leads to a peak of the elliptic flow around 5~\GeVc. For very high transverse momenta, the $v_2$ of the core contribution becomes negative. As this happens at transverse momenta where the core contribution is already very small, this has little effect on the total $v_2$ but is nevertheless interesting. We find that the negative values arise from regions close to the edge of the medium moving outward very rapidly and freezing out at early times. The contribution is stronger from fluid cells close to the tip of the ellipse. It seems plausible that this could be an effect of the curvature of the surface of the ellipse. The interface with neighboring fluid cells is smaller at the tip, leading to less slowing down due to dissipation of a given fluid cell in that region and hence larger momenta in this direction resulting in a negative $v_2$. Further investigations are necessary to demonstrate this explicitly.

The resulting $v_2$ exhibit a substantial flow contribution of thermalized \Jpsi~out to about 8-10~\GeVc~. This suggests that the flow observed in data even at these large transverse momenta may not be due to the effect of the path length difference, as commonly assumed for the high \pT region, but still is of hydrodynamic origin coming from the anisotropy of the fluid cells and their motion.

Comparison to ALICE data at central and forward rapidities shows good agreement for central collisions. For semi-central events the range in \pT of sizable $v_2$ values is reasonably well reproduced, while the model produces about twice the maximum flow amplitude.

\begin{figure}[htbp]
\centering
\includegraphics[width=.48\textwidth]{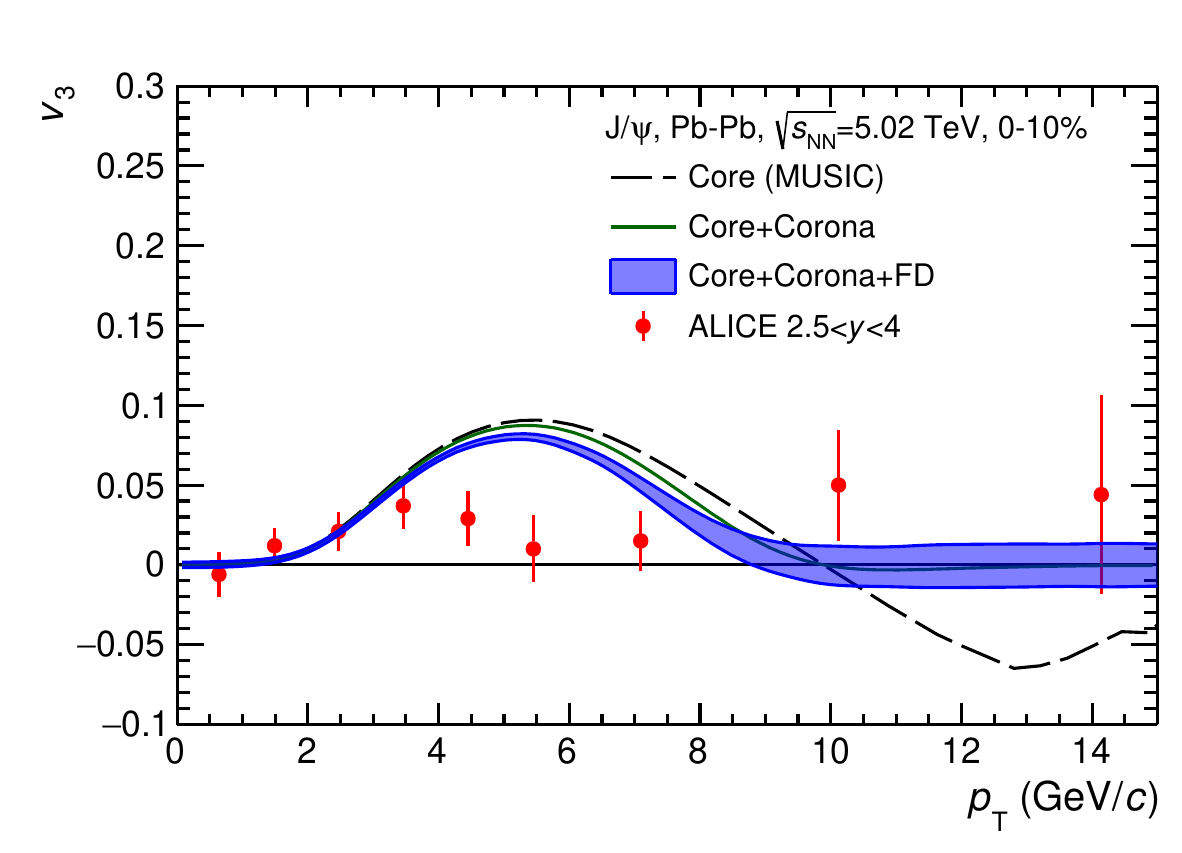}
\includegraphics[width=.48\textwidth]{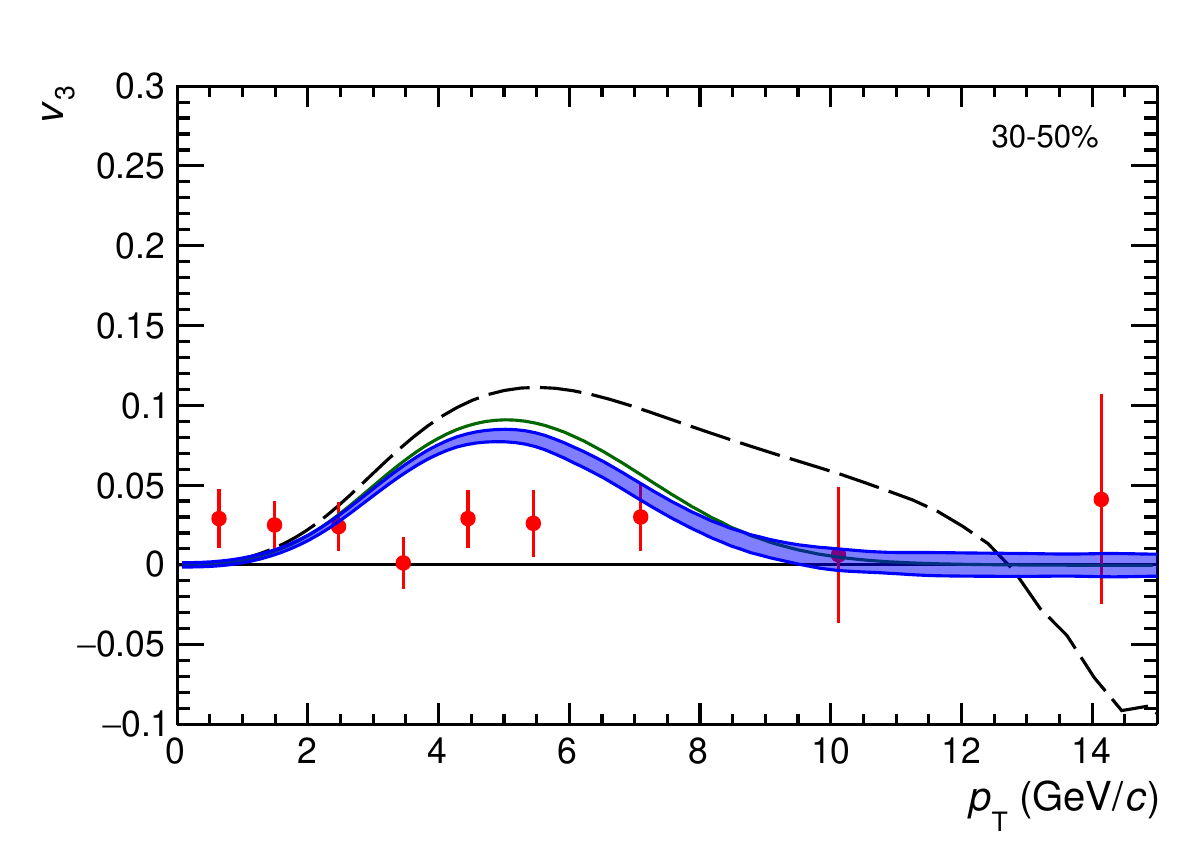}
\caption{The \, $v_3$ coefficients of the SHMc model combined with the 2+1D MUSIC calculations. The model results are compared with data from the ALICE Collaboration at forward rapidity for central (left) and semi-central (right) \PbPb collisions~\cite{ALICE:2020pvw}.\label{fig:v3}}
\end{figure}

The $v_3$ coefficients exhibit a similar behaviour for both central and more peripheral events. Similar to the $v_2$, the contribution from the core gives a substantial contribution up to $8-9~\GeVc$, which is in line with measurements. The triangular flow coefficients from our model peak at values around 0.08 for transverse momenta of 5-6 \GeVc. These peak values are somewhat larger than the experimental observations. For comparison,  the $v_2$ flow coefficients measured by ALICE in the forward direction are larger by roughly a factor of two. 

\section{Matching hydrodynamics to blast wave parametrization and resulting open charm spectra}
\label{sec::BWMatching}

To calculate the momentum distributions of open charm hadrons in the SHMc, the feed down of a large number of charm hadron states needs to be taken into account. In the SHMc 55 charmed mesons and 74 charmed (anti) baryons with their decay branching fractions are included. They are not usually included in the hydrodynamics codes, where we included the charmonia considered in the previous section explicitly. The inclusion of as complete a mass spectrum as possible is very important: Strong decays quadruple the D meson yields, for the \Lambdac the increase is by a factor 5 \cite{Andronic:2021erx}. To incorporate this the \FastReso \cite{FastReso} framework provides the necessary capability including the decay kinematics of the open charm hadrons. The input to the calculation is provided by a blast wave model \cite{Schnedermann:1993ws,Florkowski:2010zz,Andronic:2019wva}.

This model uses a power law dependence of velocity of radius $\beta(r)=\beta_{\rm max} (r/r_{\rm max})^n$ with a maximum transverse medium velocity $\beta_{\rm max}$ and an exponent $n$ of the radial dependence as input parameters. While we already employed this ansatz in our previous publications \cite{Andronic:2019wva,Andronic:2021erx}, the matching of the blast wave parameters to the hydrodynamic calculation was done differently here. Since we have now at our disposal the \Jpsi spectrum from full hydrodynamics as shown in the previous section, we can find an optimum matching of the parametrization to full hydrodynamics.

In general, the production of particles from a part of the freeze out hypersurface depends on the mass of the particles, as given in equation \ref{eq::CF}. The freeze out hypersurface is in fact quite different in the full hydrodynamic model when compared to schematic hypersurfaces used in the blast wave approach. In particular it also has surface like contributions with space like normal vectors. Thus, the blast wave parameters which best reproduce the hydrodynamic model depend on the particle mass. However, in the limit of large masses, this matching becomes easier again. Dividing eq.\ \ref{eq::CF} by the particle energy, we get:

\begin{equation}
    \frac{\drvd N}{\drvd^3 p} = g_i \int_\Sigma f(u^\mu p_\mu) \frac{1}{\gamma} u^\mu \drvd^3 \Sigma_\mu ~. \label{eq::VolumeEquivalent}
\end{equation}

In the limit of large masses, the particle velocity is approximately equal to the medium velocity as thermal velocity fluctuations are suppressed. Therefore,  we can use $u$ and $\gamma$ corresponding to the medium. Then, $\frac{1}{\gamma} u^\mu \drvd^3 \Sigma_\mu$ is independent of the particle species. If the freeze out hypersurface is taken at constant time ($\Sigma_\mu=(\Sigma_0,\vec{0})$), $\frac{1}{\gamma} u^\mu \drvd^3 \Sigma_\mu$ is simply the volume of the fluid cell. This volume can be used as a weight to take into account how fast the medium flows through the hypersurface element and thereby how much it contributes to the spectrum of the (large mass) particle. The velocities of each hypersurface element are multiplied with it's volume as a weight to find the best match of a blast wave parametrization to the full hydrodynamic result for charmed hadrons. 

\begin{figure}[htbp]
\centering
\includegraphics[width=.68\textwidth]{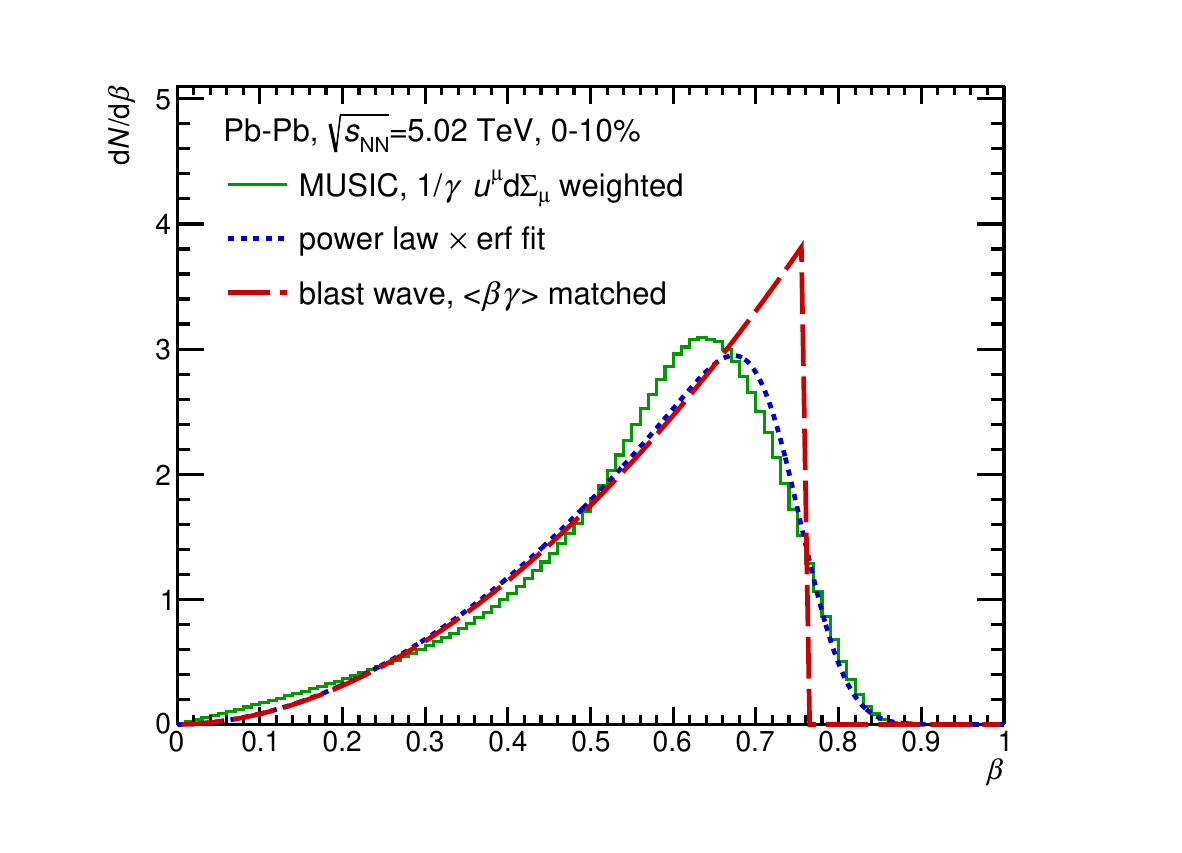}
\caption{Comparison of velocity distribution of the freeze out hypersurface elements obtained from MUSIC (green line) to a matched blast wave distribution with a smoothed edge (short dashed blue line) and a velocity matched blast wave distribution (red dashed line). \label{fig:dNdbetaMatching}}
\end{figure}

In contrast to the previous method of extraction, this ignores the spatial position of the individual fluid cells and only uses their velocity distribution. The effort here is thus to optimally reproduce the resulting charmed hadron (\Jpsi) momentum distribution of the hydrodynamic code with the blast wave model rather than reproducing the $R-\beta$ distribution. In general, there is no need to equate the radius parameter $r$ in the blast wave model with the radius in the hydrodynamic simulation. For a radius-independent freeze out time, the velocity distribution of the blast wave model has the form of a power law $\drvd N/\drvd \beta\sim \beta^\alpha$, with an abrupt drop to $0$ at $\beta_{\rm max}$. Figure \ref{fig:dNdbetaMatching} shows the volume weighted distribution of velocities of the freeze out hypersurface elements from MUSIC. Obviously there is no sharp drop but rather a smooth fall off at large $\beta$. This is due to the generally broad distribution of hypersurface elements as shown in Fig.~\ref{fig:RTau} and also event-by-event fluctuations. The matching to the MUSIC distribution is done in two steps. First the distribution from MUSIC is fitted with a power law multiplied by an error function at large $\beta$. As shown in Fig. \ref{fig:dNdbetaMatching} this distribution (blue short dashed) curve represents a very close match to the MUSIC output. This fit yields the power law parameter of the blast wave parametrization, $n=2/(\alpha+1)$. Next, using the power law exponent $n$, the $\beta_{\rm max}$ of the blast wave parametrization is fixed such that its $\langle \beta\gamma \rangle$ is equal to the one from MUSIC. This is represented by the long dashed red line in Fig. \ref{fig:dNdbetaMatching}. The resulting optimal blast wave parameters are $n$ = 0.69 and $\beta_{max}$ = 0.76. Using these parameters and the blast wave parametrization yields a momentum distribution for the \Jpsi very similar to that of the freeze out from the hydrodynamic simulations themselves (see Fig. \ref{fig:RTau} right) verifying that the current procedure produces an optimum match. Comparing these to the blast wave parameters using in \cite{Andronic:2019wva,Andronic:2021erx}, $n$ = 0.85 and $\beta_{max}$ = 0.62, the former is about 20 \% smaller and the latter 20 \% larger than the optimum match. While the effect of the increase in $\beta_{max}$ is more significant for the spectrum, both changes go into the same direction of making the spectra harder (such as the distributions shown in Figs. \ref{fig:Spectra}, \ref{fig:psi2S}) as compared to what is shown in \cite{Andronic:2019wva}.

As a further cross check we evaluate the equivalent freeze out volume represented by the hypersurface elements in MUSIC. The weights $\frac{1}{\gamma} u^\mu \drvd^3 \Sigma_\mu$ give the volume relevant for particle production in the collision center-of-momentum frame. When integrating over the momenta to get the total yield, the Lorentz contraction of this volume also needs to be considered, giving another factor of $\gamma$. Summing up the contributions $u^\mu \drvd^3 \Sigma_\mu$ thus gives an estimate for the effective volume for particle production of about 5000 $\fm^3$ very close to the result from the SHM fits \cite{Andronic:2021erx}.

\begin{figure}[htbp]
\centering
\includegraphics[width=.48\textwidth]{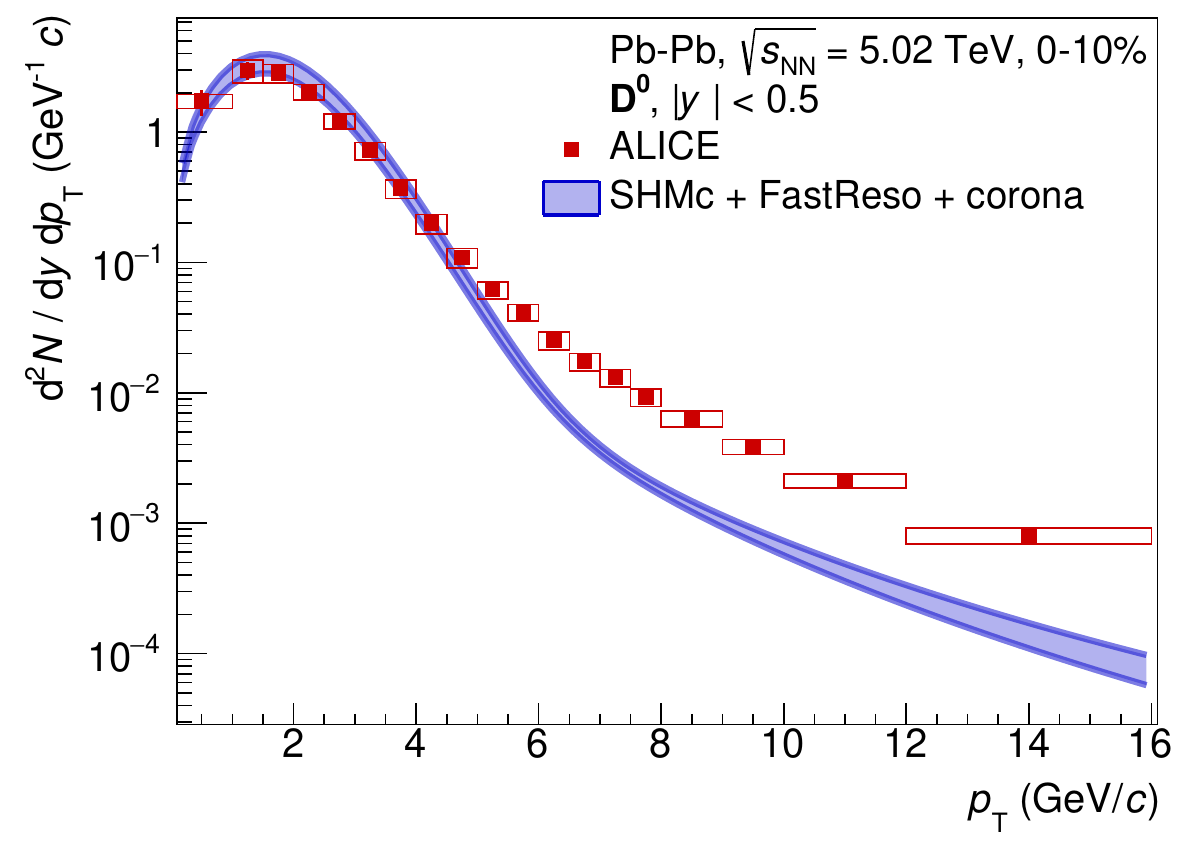}
\includegraphics[width=.48\textwidth]{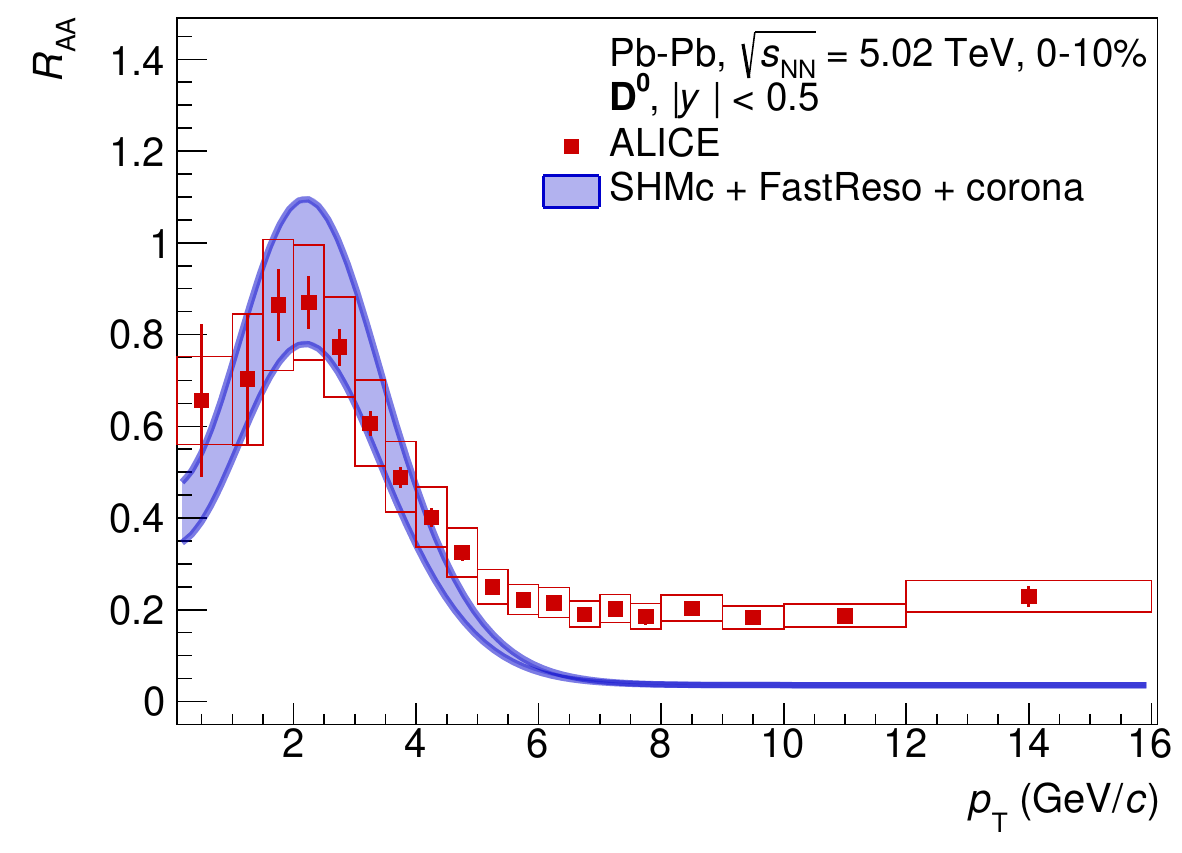}
\caption{Left: Prompt D meson spectra for central \PbPb collisions including feed down from strong decays of higher mass charmed hadrons using the optimized blast wave parametrization and \FastReso. They are compared to ALICE data \cite{ALICE:2021rxa}. Right: The same spectra normalized to those from pp collisions scaled with the number of binary collisions.\label{fig:DMesonSpectra}}
\end{figure}

The blast wave parameters were used as the input for a calculation of the \Dzero transverse momentum spectrum, considering the core contribution with feed down from strong decays as in \cite{Andronic:2021erx} and corona contribution based on the measured spectrum from pp collisions. The resulting \pT distributions are shown in Figure \ref{fig:DMesonSpectra} in comparison with data from the ALICE Collaboration. The uncertainties are determined as above for the \Jpsi spectra. Also shown in the right panel are the corresponding distributions normalized to pp collisions scaled with the number of binary collisions, i.e. \RAA. For \RAA the uncertainties introduced by the measured spectrum in pp collisions used for the corona part cancel. It can be seen that the contribution from the thermalized bulk of the medium is dominant at momenta below a few times the particle mass, about 5 GeV/$c$. This region is in very good agreement with the model. At higher momenta there is a deficit in the model. The additional contribution could be from c-quarks not fully thermalized in the medium that should show up at high \pT.

\begin{figure}[htbp]
\centering
\includegraphics[width=.48\textwidth]{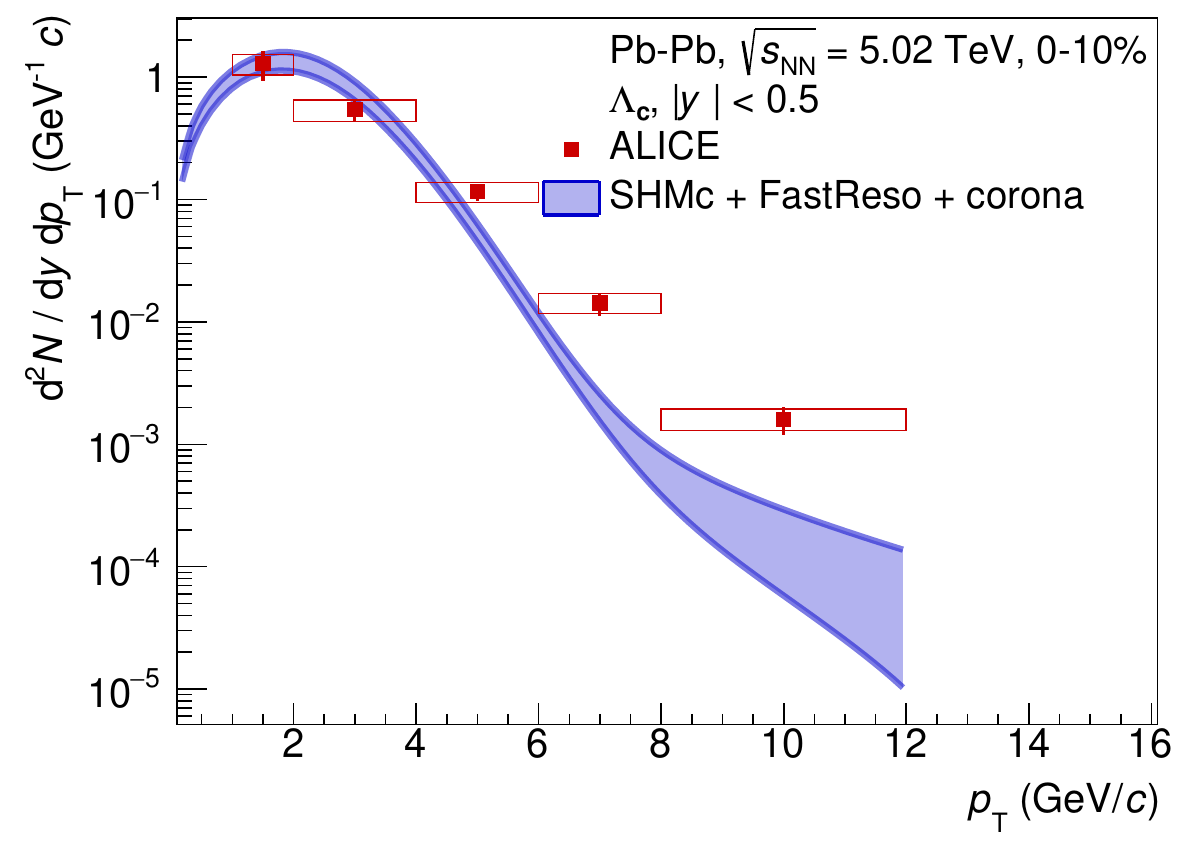}
\includegraphics[width=.48\textwidth]{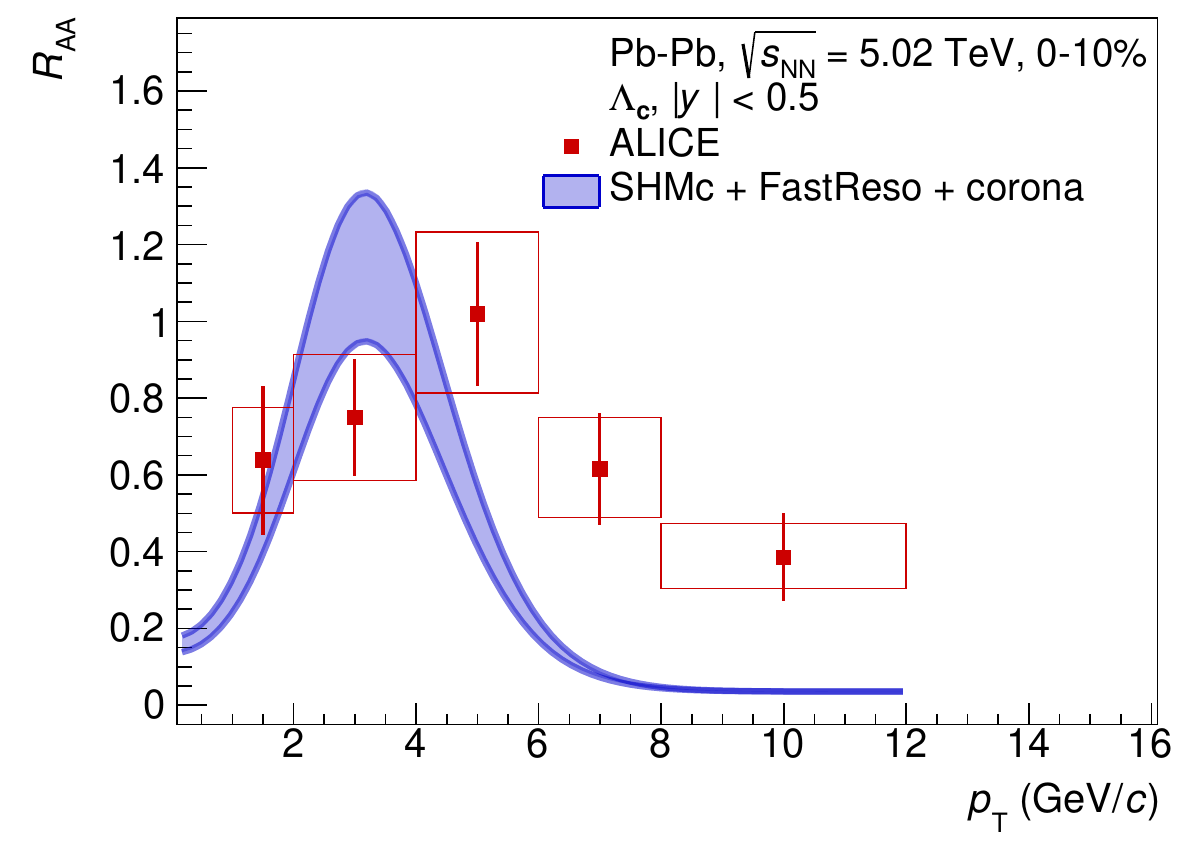}
\caption{Left: Spectra of \Lambdac for central \PbPb collisions including feed down from strong decays of higher mass charmed baryons using the optimized blast wave parametrization and \FastReso. For this calculation additional undiscovered charm baryon states were considered (see \cite{Andronic:2021erx}). Also shown are ALICE data \cite{ALICE:2021bib}. Right: The same spectra normalized to those from pp collisions scaled with the number of binary collisions. \label{fig:Lambdac}}
\end{figure}

The transverse momentum distributions of all other charmed hadrons can be computed in an analogue way. We show here in Figure \ref{fig:Lambdac} the \pT distribution calculated for \Lambdac together with the corresponding \RAA distribution. Displayed are results from the SHMc with an enhanced charmed baryon scenario as outlined in \cite{Andronic:2021erx}. The motivation is the observation by the ALICE collaboration of a very large fragmentation fraction of charm into the \Lambdac baryon in pp collisions at the LHC \cite{ALICE:2021dhb}. One explanation, proposed by He and Rapp \cite{He:2019tik}, was based on the observation that the charmed baryon spectrum from the relativistic quark model contains many more charmed baryon states than established experimentally. This is also the case for the charmed baryon spectrum from lattice QCD \cite{Bazavov:2014yba}. The hypothesis was made that these states exist but were not yet discovered experimentally. This could be due to a large width and therefore significant overlap. Including these theoretically predicted states into their statistical model, He and Rapp found that they could reproduce the enhanced fragmentation into \Lambdac. In the same spirit, we have enhanced the contribution of charmed baryons by the equivalent amount in our SHMc calculations and find that the \Lambdac yield in \PbPb collisions is about doubled while the D meson sector is practically unchanged, for a correspondingly larger charm cross section (see \cite{Andronic:2021erx} where both the regular and the enhanced charm baryon results are shown). 

\begin{figure}[htbp]
\centering
\includegraphics[width=.52\textwidth]{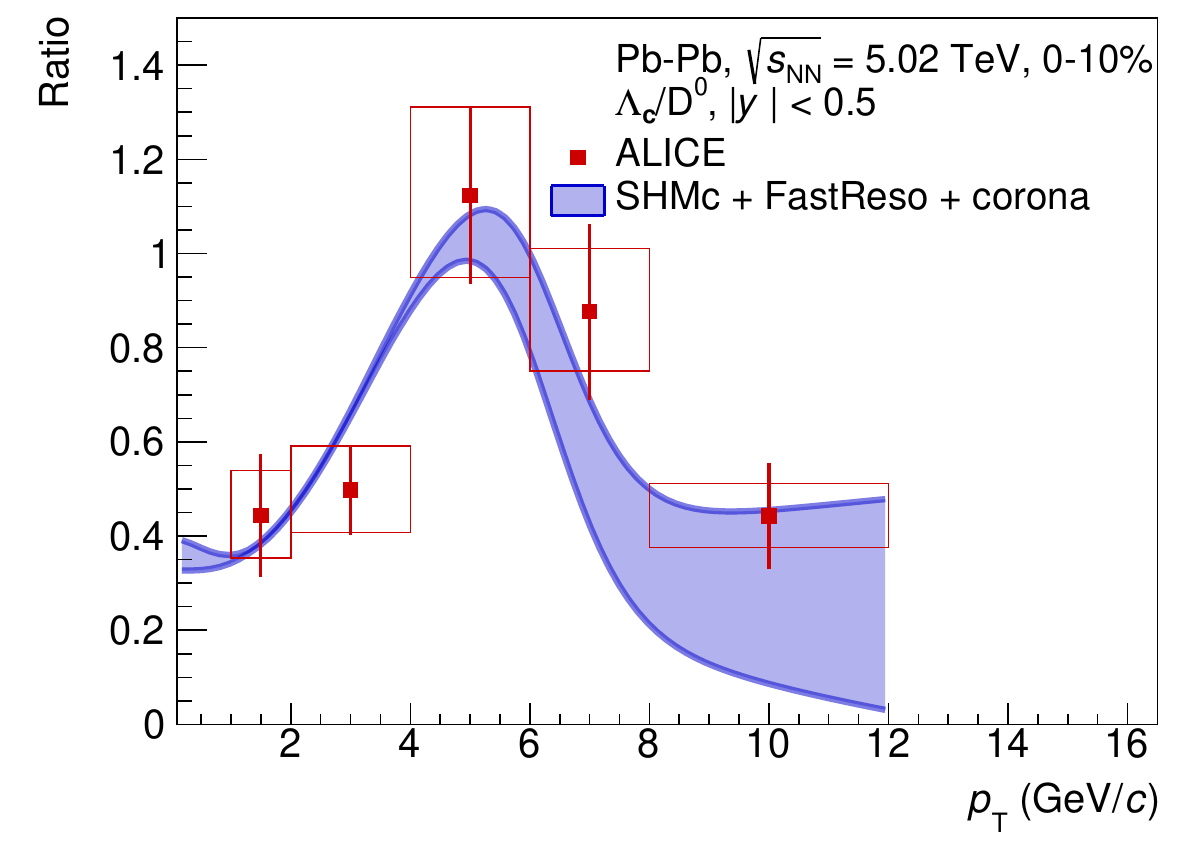}
\caption{Ratio of \Lambdac to \Dzero yields as function of transverse momentum for central \PbPb collisions as shown separately in Figs: \ref{fig:DMesonSpectra}, \ref{fig:Lambdac} compared to ALICE data \cite{ALICE:2021bib,ALICE:2021rxa}. \label{fig:LamdactoD}}
\end{figure}

The resulting \Lambdac spectral distribution (Fig. \ref{fig:Lambdac}) is compared to recent data from ALICE and a similar picture emerges as for charmonia and D mesons: The yield at low and moderate \pT and therefore the bulk yield are well reproduced by the current calculations, in particular considering the uncertainties. At high \pT there is again a deficit in the model results. We also show in Figure \ref{fig:LamdactoD} the ratio of \Lambdac to D meson \pT spectra. The experimental data from ALICE are very well reproduced. In particular the maximum in this baryon to meson ratio comes out naturally from the hydrodynamic description combined with statistical hadronization without invoking an additional coalescence mechanism. 

\section{Estimating systematic effects of charm space time distribution in the medium}
\label{sec:charmspacetime}

In the SHMc approach to charm hadronization we assume that the charm quarks, at hadronization, are thermalized in terms of the momentum distributions, i.e. the momentum distribution is determined by a (common) freeze out temperature and the fluid expansion at this instance.
This is supported by observables of D mesons compared to transport models where a range is determined for the spatial diffusion coefficient
$1.5 < 2 \pi D_s < 4.5$ at the pseudo critical temperature \cite{ALICE:2021rxa}. From this a kinetic equilibration time $\tau_{kin}$ = 2.5 - 7.6 fm/$c$ can be estimated. Recently, for the first time spatial diffusion coefficients were derived in lattice QCD calculations with light dynamical quarks \cite{Altenkort:2023oms}. In the region between 1 and 2 times the pseudo critical temperature the values were found to be significantly smaller than in previous quenched lQCD calculations, implying very fast hydrodynamization of charm quarks within 1-2 fm/$c$. This rapid thermalization does however not imply that the spatial distribution of charm quarks at hadronization is the same as that of light quarks and gluons. 

While the initial energy density is distributed roughly following the areal density of participants $N_{part}$, the charm quark density follows the areal density of binary collisions $N_{coll}$ which is more compact. Therefore, at early times the outermost region of the nuclear overlap region is less densely populated with charm quarks as compared to energy, or equivalently, light quarks and gluons.

We have no experimental observable at present that gives any information about the spatial distribution of charm quarks at hadronization. In fact, since the momentum equilibration takes some time (several collisions), it is plausible that the expansion of charm quarks lags somewhat behind that of the overall medium and even at late times the ou\-termost regions of the freeze out hypersurface may not be equally populated with charm quarks. In order to give an indication how charmed hadron momentum distributions would be affected, we present in the following two very schematic estimates basically removing  contributions from (i) early times or (ii) large radii at late times.

\begin{figure}[htbp]
\centering
\includegraphics[width=.5\textwidth]{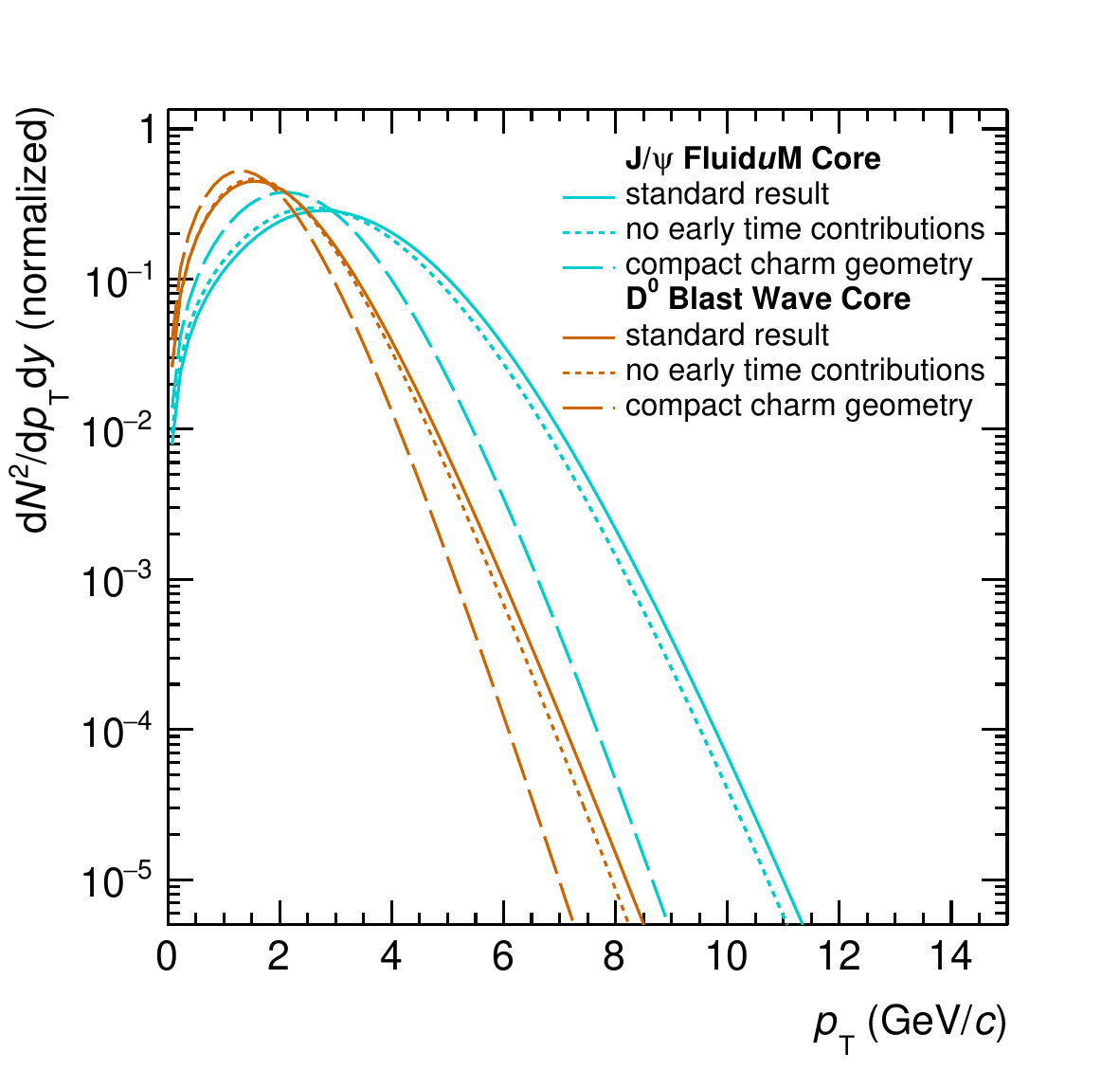}
\caption{Comparison of core components of D meson and \Jpsi spectra with the default freeze out, to those calculated removing the early freeze-out (short dashed lines) and to those removing, in the blast wave parametrization, the outermost radii (long dashed lines). The spectra are normalized to the same integral for easier comparison. For the \Dzero mesons, only the direct contribution is shown. \label{fig:CharmDiffusionComparison}}
\end{figure}

The distribution of hypersurface elements at freeze out is shown in Fig. \ref{fig:RTau} and it is visible that the outermost regions of the nuclear overlap freeze out within the first few fm/$c$. These are regions that, immediately after the collision, are underpopulated in charm as compared to energy. And at these early times it is also questionable that charm quarks are equilibrated in terms of their momenta. To test the effect on D meson spectra, we remove contributions from the hypersurface elements corresponding to the earliest 5 \% of the effective volume at freeze out. This happens to correspond to radii where the density of $N_{coll}$ is less than half the density of $N_{part}$. Since in statistical hadronization \Jpsi mesons are formed from uncorrelated charm and anticharm quarks, we cut off the earliest 15 \% for this study (considering the density of $N_{coll}^2$ vs density of $N_{part}$). 
This corresponds to removing the hypersurface elements located within the first 4.5 fm/$c$ (see Fig. \ref{fig:RTau}) for D meson production and 6.7 fm/$c$ for \Jpsi production. We recalculate blast wave parameters matching the remaining distribution and obtain for D mesons $n=0.647$ and $\beta_{\rm max}=0.74$. In fact some of the elements with the largest velocity were removed by this cut. The resulting spectra for D mesons and \Jpsi are shown in Fig. \ref{fig:CharmDiffusionComparison} as the short dashed lines in comparison to the full spectra. Note that for this discussion only the core component is shown. Once can see a visible but small effect in terms of softening of the \pT spectra. 

To test the scenario that even at late times the charm quark spatial distribution could be more compact than the overall fireball, we resort to the blast wave distribution of velocities matched to the MUSIC freeze out hypersurface and remove the outermost 1 fm. This simply implies $\beta_{max}$ = 0.69 instead 0.76. The corresponding spectra are also shown in Fig. \ref{fig:CharmDiffusionComparison} as the long dashed lines. Now  a considerable softening of the distributions is observed. The most probable \pT shifts downward for D mesons by 0.3 GeV/$c$ and for \Jpsi by nearly 1 GeV/$c$.

While the two scenarios discussed are very schematic, it is obvious that they both go in the direction of obtaining an even better match to experimental data for \Jpsi, where we currently see a deviation, while much less affecting the D meson spectrum, where we already see a good match between model and data. It is also obvious that a slightly more compact final state has a bigger effect than removal of early freeze out. 

\section{Summary and Outlook}

In this paper we have explored the possibilities to describe quantitatively and with essentially no free parameters the transverse momentum spectra and anisotropic flow distributions for charmonia and charmed hadrons produced in \PbPb collisions and measured  with the ALICE detector at the CERN Large Hadron Collider (LHC). The basis for this work is the realization that the SHM originally developed to describe yields of hadrons composed of (u,d,s) quarks can be expanded into the charm sector with no free additional parameters by treating charm quarks as 'impurities' in the hot and dense fireball with the charm quark number determined by the  measured total open charm cross section in \PbPb or \AuAu collisions. Furthermore, the transverse dynamics in the charm sector is more straightforward to implement than in the light quark sector  as we have evidence that all hadrons with heavy quarks are formed at or very near the QCD phase boundary. 

The specific focus of the present paper is indeed the transverse dynamics in the charm quark sector which is further developed by coupling the SHMc to the computer codes  MUSIC and \Fluidum. For charmonia and hadrons with one charm quark a rather quantitative description is obtained in the low \pT sector for spectra of D-mesons, \Lambdac baryons and charmonia. The observed wide distribution in \pT of anisotropic flow coefficients v$_2$ and v$_3$ for charmonia is also well reproduced, but their magnitude is generally somewhat over predicted.

This finding may be connected to a difference in spatial distribution  between light and charmed hadrons due to a different diffusion of light and heavy quarks in the hot fireball. To make progress in this area two scenarios are explored by modifying the distribution of charm quarks at late and early times. The first results are promising but further development is needed to make these studies quantitative. 

On the theory side, to make progress requires more thorough modelling for instance in the direction employed in~\cite{Capellino:2023cxe}. On the experimental side, a measurement of Hanbury Brown-Twiss correlations between D mesons would give experimental evidence about the charm quark spatial distribution.

Finally, a completely open field awaits us in the area of hadrons with 2 or 3 charm quarks. These will be experimentally explored in LHC Run 3 and Run 4 with the upgraded ALICE apparatus and ultimately with the proposed new ALICE 3 detector~\cite{ALICE:2022wwr}. It will be very exciting to test SHMc predictions in the multi-charm sector and to explore hadronization and deconfinement scenarios there.

\acknowledgments

We would like to thank A.~Kirchner, C.~Shen and B.~Schenke for their support concerning \Fluidum~and MUSIC. We acknowledge inspiring discussions with S.~Floerchinger, B.~Friman, A.~Mazeliauskas and K.~Redlich. This work was performed within and supported by the DFG Collaborative Research Centre "SFB 1225 (ISOQUANT)". V. V. gratefully acknowledges funding from the Knut and Alice Wallenberg Foundation (the CLASH project).


\bibliographystyle{JHEP}
\bibliography{bibliography.bib}


\end{document}